\documentstyle[preprint,eqsecnum,aps,tighten,epsbox]{revtex}

\begin{document}
\draft
\title{An Oscillator Neural Network Retrieving Sparsely Coded Phase Patterns}
\author{Toshio Aoyagi and  Masaki Nomura}
\address{Department of Applied Mathematics and Physics, Graduate School of 
Informatics, Kyoto University, Kyoto, 606-8501, Japan}
\date{November 6, 1988}
\maketitle
\begin{abstract}
Little is known theoretically about the associative memory capabilities of neural 
networks in which information is encoded not only in the mean firing rate 
but also in the timing of firings. 
Particularly, in the case that the fraction of active neurons involved
in memorizing patterns becomes small,
it is biologically important to consider the timings of firings and to study how such
consideration influences storage capacities and quality of recalled patterns.
For this purpose, we propose a simple extended model of oscillator neural networks 
to allow for expression of non-firing state.
Analyzing both equilibrium states and dynamical properties in recalling processes,
we find that the system possesses good associative memory.
\end{abstract}

\narrowtext

Since several recent experiments suggest that the temporal coherence of neuronal 
activity, such as the synchronization of pulses, plays a significant role in real 
neuronal systems\cite{gray},
a great number of authors have proposed and studied many theoretical models of 
neural networks in attempt to understand the essential dynamics of the these systems
\cite{malsburg,Sompolinsky,terman}.
Among these models, the neural networks of phase oscillators provide a useful 
framework for modeling and analyzing such temporal behavior in neuronal systems.
The main reason for the usefulness of these models is their mathematical 
tractability, which has allowed us to obtain many important analytic results. 
Another reason is that, using a certain mathematical technique, 
the complex dynamics of coupled oscillatory neuronal systems under suitable 
conditions can be reduced to the dynamics of phase oscillator 
\cite{erementrout,kuramotobook}. 
Since the relation between real systems and these models is theoretically clear,
it is expected that their analysis will shed light on the role of oscillatory 
behavior in real neuronal systems.
In particular, the properties of oscillator neural networks with regard to 
associative memory have been studied recently by many authors
\cite{Abbott,cock,Noest,aoyagik,aoyagi2}.

With regard to the above mentioned framework, 
it should be noted that all neurons are assumed to 
exhibit periodic firing states at all times. 
However, such an assumption is not realistic from a biological viewpoint.
This is because, in real systems, whether or not a neuron is firing usually 
depends on the situation, in particular the pattern it is presently recalling.
In fact, it is well known that only a small fraction of neurons are active at a
given time in the central nervous system.
Situations in which the level of activity is very low are often termed
{\it sparse coding}\cite{willshow}.
In the case of standard binary neurons, it is found that the storage capacity 
for such sparse coding diverges as $-1/a \ln a$, where $a$ is the fraction of 
active neurons\cite{tsodyks,vicente,buhmann,okada-sparse}. 
This is the optimal asymptotic form\cite{gardner}.
It is biologically plausible that this theoretical optimal bound is achieved even 
for a simple Hebbian learning rule.

In light of the above considerations, in order to construct more realistic models,
it is natural to consider the encoding of information in both the mean firing rate 
and the relative timing of neuronal spikes.
However, little is theoretically known about the properties of associative memory 
for sparse coded patterns in such systems.
For example, one of the interesting questions is whether the storage capacity
in such systems, as in the standard binary model, diverges as the activity level 
$a$ decreases.
Another essential point is the quality of the recalled pattern, particularly 
concerning the timing of the spikes.
To clarify these points, we first need to extend the phase oscillator model 
to allow for expression of the non-firing state\cite{Aoyagi}.
In this paper, we propose and examine a simple extended model as a first step 
toward the theoretical study of neuronal systems in which the timing of spikes can 
carry information.

Let us start with a brief review of the theoretical basis of the phase oscillator model.
It is well known that in coupled oscillatory neuronal systems, under suitable 
conditions, the original dynamics can be reduced theoretically to a simpler phase 
dynamics.
The state of the $i$th neuronal oscillatory system can be then
characterized by a single phase variable $\phi_i$ representing the timing of the 
neuronal firings.
The typical dynamics of oscillator neural networks are described by equations
of the form
\cite{erementrout,kuramotobook,hoppensteadt}.
\begin{equation}
{d\phi_i\over dt}=\omega_i+\sum_{j=1}^N J_{ij}\sin(\phi_j -\phi_i+\beta_{ij}).
\label{phasemodel}
\end{equation}
Here, $J_{ij}$ and $\beta_{ij}$ are parameters representing the effect of the 
interaction.
For simplicity, we assume that all natural frequencies $\omega_i$ are equal to
some fixed value $\omega_0$.
We can then eliminate $\omega_0$ by applying the transformation
$\phi_i\rightarrow\phi_i+\omega_0 t$.
Using the complex representation $W_i=\exp(i\phi_i)$ and 
$C_{ij}=J_{ij}\exp(i\beta_{ij})$ in (\ref{phasemodel}),
it is easily found that all neurons relax toward their stable equilibrium states,
in which the relation $W_i=h_i/|h_i|$ ($h_i=\sum_{j=1}^N C_{ij}W_j $) is satisfied.
Following this line of reasoning, as a synchronous update version of the oscillator 
neural network we can consider the alternative discrete form\cite{Noest,aoyagik,aoyagi2},
\begin{equation}
W_i(t+1)={h_i(t)\over |h_i(t)|}
\hskip 0.5cm h_i(t)=\sum_{j=1}^N C_{ij}W_j(t) .
\label{ampmodel}
\end{equation}

Now we will attempt to construct an extended model of the oscillator neural 
networks to retrieve sparsely coded phase patterns.
In equation (\ref{ampmodel}), the complex quantity $h_i$ can be regarded as the 
local field produced by all other neurons.
We should remark that the phase of this field, $h_i$, determines the timing of the 
$i$th neuron at the next time step, while the amplitude $|h_i|$ has no effect on 
the retrieval dynamics(\ref{ampmodel}).
It seems that the amplitude can be thought of as the strength of the local field 
with regard to emitting spikes. 
Pursuing this idea, as a natural extension of the original model we stipulate that 
the system does not fire and stays in the resting state if the amplitude is smaller
than a certain value.
Therefore, we consider a network of $N$ oscillators whose dynamics are governed by
\begin{equation}
W_i(t+1)=f(|h_i(t)|) {h_i(t)\over |h_i(t)|}
\hskip 0.5cm h_i(t)=\sum_{j=1}^N C_{ij}W_j(t) .
\label{sparsemodel}
\end{equation}
In this paper, we assume that $f(x)=\Theta(x-H)$, where the real variable $H$ is 
a threshold parameter and $\Theta(x)$ is the unit step function; $\Theta(x)=1$ 
for $x \geq 0$ and $0$ otherwise.
Accordingly, the amplitude $|W_i^t|$ assumes a value of either 1 or 0, representing
the state of the $i$th neuron as firing or non-firing.
Consequently, the neuron can emit spikes when the amplitude of the local field 
$h_i(t)$ is greater than the threshold parameter $H$.

Now, let us define a set of $P$ patterns to be memorized as 
$\xi^\mu_i=A_i^\mu\exp(i\theta_i^\mu)$ ($\mu=1,2,\dots,P$), where $\theta_i^\mu$ 
and $A_i^\mu$ represent the phase and the amplitude of the $i$th neuron in the 
$\mu$th pattern, respectively.
For simplicity, we assume that the $\theta_i^\mu$ are chosen at random from a 
uniform distribution between $0$ and $2\pi$.
The amplitudes $A_i^\mu$ are chosen independently with the probability distribution
\begin{equation}
P(A_i^\mu)=a\delta(A_i^\mu-1)+(1-a)\delta(A_i^\mu),
\end{equation}
where $a$ is the mean activity level in the patterns.
Note that, if $H=0$ and $a=1$, this model reduces to (\ref{ampmodel}).

For the synaptic efficacies, to realize the function of the associative memory, 
we adopt the generalized Hebbian rule in the form
\begin{equation}
C_{ij}={1\over aN}\sum_{\mu=1}^P \xi^\mu_i \tilde\xi^\mu_j ,
\label{coupling}
\end{equation}
where $\tilde\xi^\mu_j$ denotes the complex conjugate of $\xi^\mu_j$.
The overlap $M_\mu(t)$ between the state of the system and the pattern $\mu$ at 
time $t$ is given by
\begin{equation}
M_\mu(t)=m_\mu(t) e^{i\varphi_\mu(t)}={1\over aN}\sum_{j=1}^N \tilde\xi^\mu_j W_j(t),
\label{overlap}
\end{equation}
In practice, the rotational symmetry forces us to measure the correlation of the 
system with the pattern $\mu$ in terms of the amplitude component 
$m_\mu(t)=|M_\mu(t)|$.

Let us consider the situation in which the network is recalling the pattern 
$\xi_i^1$;that is, $m_1(t)=m(t)\sim O(1)$ and $m_\mu(t)\sim O(1/\sqrt{N}) 
(\mu\not= 1)$.
The local field $h_i(t)$ in Eq.(\ref{sparsemodel}) can then be separated as
\begin{equation}
h_i(t)=\sum_{j=1}^N  C_{ij}W_j(t)
=m_t e^{i\varphi_1(t)} \xi^1_i + z_i(t) ,
\label{localfield}
\end{equation}
where $z_i(t)$ is defined by
\begin{equation}
z_i(t)={1\over aN}\sum_{j=1}^N \sum_{\mu=2}^P \xi_i^\mu \tilde\xi^\mu_j W_j(t) .
\label{zdef}
\end{equation}
The first term in Eq.(\ref{localfield}) acts to recall the pattern, while the 
second term can be regarded as the noise arising from the other learned patterns.
The essential point in this analysis is the treatment of the second term as complex
Gaussian noise characterized by
\begin{equation}
<z_i(t)>=0, <|z_i(t)|^2>=2\sigma(t)^2 .
\label{zassump}
\end{equation}
In addition, we also assume that $\varphi_1(t)$ remains a constant, that is, 
$\varphi_1(t)=\varphi_0$.
By applying the method of statistical neurodynamics to this model under the above
assumptions\cite{shiino,amari,okada,coolen}, 
we can study the retrieval properties analytically.
As a result of such analysis we have found that the retrieval process can be 
characterized by some macroscopic order parameters, such as $m(t)$ and $\sigma(t)$.
However, in this short paper, we give only a rough explanation of this derivation 
and report only the most relevant results.  
A detailed derivation will be published elsewhere in the near future.

From Eq.(\ref{overlap}), we easily find that the overlap at time $t+1$ is given by
\begin{equation}
m(t+1)=\left<\!\left<f(|m(t)+z(t)|){m(t)+z(t) \over |m(t)+z(t)|}\right>\!\right> ,
\label{nextm}
\end{equation}
where $\left<\!\left<\cdots\right>\!\right>$ represents an average over the complex 
Gaussian $z(t)$ with mean $0$ and variance $2\sigma(t)^2$.
For the noise $z(t+1)$, in the limit $N\rightarrow\infty$ we obtain 
\begin{equation}
\begin{array}{rl}
z_i(t+1)\sim&\displaystyle\strut 
{1\over aN}\sum_{j=1}^N \sum_{\mu=2}^P \xi_i^\mu \tilde\xi^\mu_j 
f(|h_{j,\mu}(t)|){h_{j,\mu}(t) \over |h_{j,\mu}(t)|}\\
&\displaystyle\strut +z_i(t)\left(
{f^\prime(|h_{j,\mu}(t)|)\over 2}+{f(|h_{j,\mu}(t)|) \over 2|h_{j,\mu}(t)|}\right),
\end{array}
\label{nextz}
\end{equation}
where $h_{j,\mu}(t)=1/aN\sum_{k=1}^N\sum_{\nu\not=mu,1}^P
\xi_j^\nu \tilde\xi^\nu_k W_k(t)$.
We should remark that the $h_{j,\mu}(t)$ can be regarded as independent of $\xi_i^\mu$.
Using this fact, we can easily carry out various calculations for $\sigma(t+1)$.

\begin{figure}[t]
\centerline{
\epsfile{file=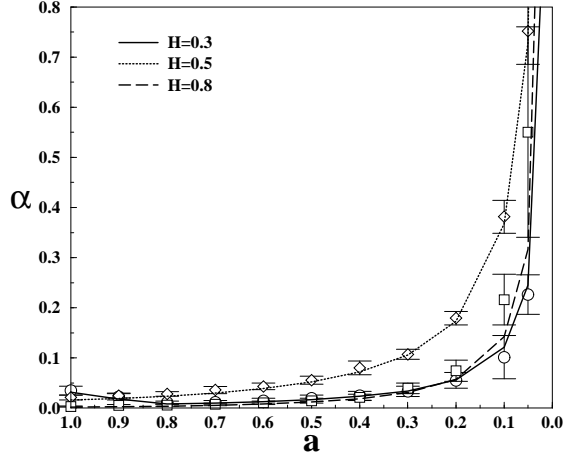,scale=0.45}}
\caption{Dependence of the storage capacity $\alpha_c$ on the mean activity level $a$
for various threshold parameters $H$.
\label{alphac}
}
\end{figure}
First, we consider the equilibrium state of this system. In the equilibrium state
we have $z_i(t)=z_i(t+1)=z_i$ $(\sigma(t)=\sigma(t+1)=\sigma)$, $m(t)=m(t+1)=m$.
Using Eqs.(\ref{nextm}) and (\ref{nextz}), we finally find that the values of $m$ 
and $\sigma$ are determined by the equations
\begin{equation}
\left\{\begin{array}{rl}
m=& \displaystyle\strut M(m,\sigma) \\
\sigma^2=& \displaystyle\strut 
{\alpha Q(m,\sigma) \over 2\left(1-G(m,\sigma)\right)^2} ,
\end{array}\right. 
\label{stateeq}
\end{equation}
with
\begin{equation}
\begin{array}{rl}
M(m,\sigma)=& \displaystyle\strut
\!\!\int_H^\infty \!\!\!dr\int_0^\pi \!\!\!d\theta {r\cos\theta\over\pi\sigma^2}
\exp\left(-{m^2+r^2-2mr\cos\theta\over 2\sigma^2}\right) \\
Q(m,\sigma)=&\displaystyle\strut 
(1-a)\exp\left(-{H^2\over 2\sigma^2}\right) \\
&\hskip -1cm\displaystyle\strut
+\!\!\int_H^\infty \!\!dr\int_0^\pi \!\!d\theta {ar\over\pi\sigma^2}
\exp\left(-{m^2+r^2-2mr\cos\theta\over 2\sigma^2}\right)\\
G(m,\sigma)=& \displaystyle\strut 
{aH\over 2\pi\sigma^2}\int_0^\pi d\theta
\exp\left(-{m^2+H^2-2mH\cos\theta\over 2\sigma^2}\right)\\
&\hskip -1cm\displaystyle\strut 
+\!\!\int_H^\infty \!\!dr\int_0^\pi \!\!d\theta  {a\over 2\pi\sigma^2}
\exp\left(-{m^2+r^2-2mr\cos\theta\over 2\sigma^2}\right) \\
&\hskip -1cm\displaystyle\strut 
+{(1-a)\over 2\sigma^2}
\left[H\exp\left(-{H^2\over 2\sigma^2}\right)
+\int_H^\infty dr \exp\left(-{r^2\over 2\sigma^2}\right)\right] .
\end{array}
\label{qgdef}
\end{equation}
Here, the parameter $\alpha$ is defined by $\alpha=P/N$.
If $\alpha$ is smaller than the maximum storage capacity $\alpha_c$, 
there exists a solution of Eq.(\ref{stateeq}) for which $m\not= 0$.
This solution corresponds to a retrieval state.
It disappears at $\alpha_c$, where the overlap $m$ drops from the finite 
value $m_c$ to zero.
In Fig.(\ref{alphac}), the storage capacity $\alpha_c$ obtained from numerical 
solutions of Eq.(\ref{stateeq}) is plotted as a function of the mean activity 
level of the memorized patterns $a$ for various threshold parameters $H$.
As shown there, the storage capacity increases as the activity level decreases. 
Moreover, we have numerically confirmed that, like the binary model, 
in the limit $a\rightarrow 0$ the storage capacity diverges as $-1/a\ln a$. 

Next, we wish to discuss the performance of the system with regard to associative 
memory, that is, its associative ability to dynamically correct a noisy pattern.
This measure of the model's performance roughly corresponds to the size of the 
basin of attraction, which is obtained from analysis of the dynamics of the 
recalling process.
Following a previous work\cite{aoyagi2}, 
we can derived the recursion equations describing the retrieval dynamics.
In order to obtain the correct results in the case of oscillator neural networks, 
it is essential to take into account the fact that the noise $z_i(t)$ is temporally 
correlated.
In the $n$th order approximation, the correlation of the noise $z_i(t)$ up to 
$z_i(t-n+1)$ is correctly estimated. 

At first order ($n=1$), ignoring the temporal correlation of the noise $z_i(t)$,
we obtain 
\begin{equation}
\left\{\begin{array}{rl}
m(t+1)=& \displaystyle\strut M(m(t),\sigma(t)) \\
\sigma(t+1)^2=& \displaystyle\strut 
{\alpha\over 2}Q(m(t),\sigma(t))+\sigma(t)^2 G(m(t),\sigma(t))^2\\
&+a^2\alpha G(m(t),\sigma(t))m(t)m(t+1) .
\end{array}\right. 
\label{firstdyn}
\end{equation}
This result corresponds to the Amari-Maginu theory in the case of traditional 
neural networks\cite{amari}.

At second order (n=2), taking account of the correlation between $z_i(t)$ and 
$z_i(t-1)$, we find that 
\begin{equation}
\begin{array}{rl}
m(t+1)=&M(m(t),\sigma(t))\\
\sigma(t+1)^2=& \displaystyle\strut 
{\alpha\over 2}Q(m(t),\sigma(t))+\sigma(t)^2 G(m(t),\sigma(t))^2 \\
&+\alpha G(m(t),\sigma(t))X(t+1,t) \\
&\hskip -2cm +a^2\alpha G(m(t),\sigma(t))G(m(t-1),\sigma(t-1))m(t+1)m(t-1)\\
X(t+1,t)=& \displaystyle\strut \mbox{Re}\left<\!\left< a Y_1(t)\tilde Y_1(t-1) 
+(1-a) Y_0(t)\tilde Y_0(t-1) \right>\!\right>_{z(t),z(t-1)}\\
Y_k(t)=& \displaystyle\strut f(|\delta_{k,1}m(t)+z(t)|){\delta_{k,1}m(t)+z(t) 
\over |\delta_{k,1}m(t)+z(t)|} (k=0,1),
\end{array}
\label{seconddyn}
\end{equation}
where $\left<\!\left<\cdots\right>\!\right>_{z(t),z(t-1)}$ represents an average 
over the two complex Gaussian functions $z(t)$ and $z(t-1)$ with the correlation 
$\rho(t,t-1)=\left<\!\left<z(t)\tilde z(t-1)\right>\!\right>/2\sigma(t)\sigma(t-1)$.
In addition, estimating $\left<\!\left<z(t)\tilde z(t-1)\right>\!\right>$ by 
using Eq.(\ref{nextz}), we find that $\rho(t,t-1)$ can be calculated as
\begin{equation}
\rho(t,t-1)={\alpha X(t,t-1)\over 2\sigma(t)\sigma(t-1)}+{\sigma(t-1)
G(m(t-1),\sigma(t-1))\over\sigma(t)}.
\label{seconddynrho}
\end{equation}
At the initial time, we choose the initial value of the overlap $m(0)$ with the 
condition that $\sigma(0)^2=a\alpha/2$, $X(0,-1)=0$ and $X(1,0)=a^2 m(1)m(0)$.

\begin{figure}
\centerline{
\epsfile{file=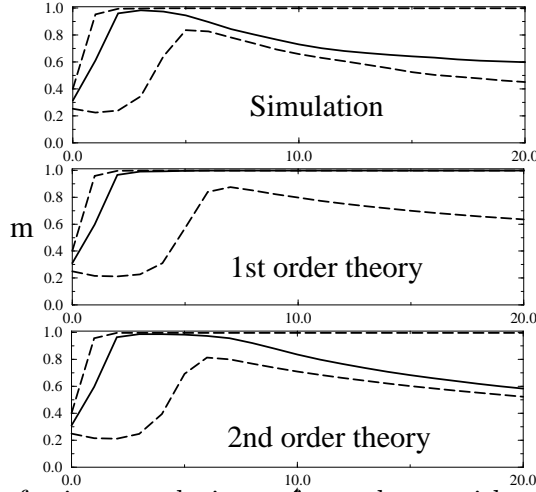,scale=0.4}}
\caption{Comparison of time evolution of overlaps with various initial conditions, 
$m=0.25,0.31,0.4$ for $\alpha=0.013$, $H=0.3$ 
and $a=0.5$ .
(a) A typical result of numerical simulations with $N=1000$.
(b) Theoretical curves at first order.
(c) Theoretical curves at second order.
\label{timeevol}
}
\end{figure}
\begin{figure}
\centerline{
\epsfile{file=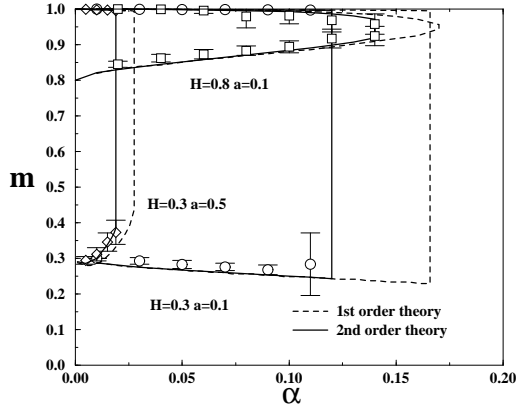,scale=0.4}}
\caption{The basin of attraction for the cases
$(H,a)=(0.3,0.5), (0.8,0.1)$ and $(0.3,0.1)$.
The solid lines and the broken lines represent the theoretical results at 
second order and at first order, respectively.
The data points indicate simulation results with $N=1000$ for $20$ trials.
\label{basin}
}
\end{figure}
In Fig.\ref{timeevol}, we show the time evolution of the overlap predicted by 
our theory and typical behavior found through numerical simulations.
As shown in this figure, the predictions obtained from the second order 
approximation are in good agreement with numerical results.
However, qualitative discrepancy between the first order theory and the numerical 
results can be found (See solid curves).
For $\alpha<\alpha_c$, only if the initial overlap is larger than a certain critical 
value $m^c_\alpha(0)$, the system generally evolves toward the retrieval state 
$m(\infty)\sim O(1)$. 
The size of the basin of attraction can be measured by this critical value.
The nature of the basin of attraction is described in Fig.\ref{basin}, which 
displays the results of theoretical analysis and numerical simulation 
as a phase diagram for various thresholds $H$ and activity levels $a$. 
The upper parts and lower parts of the theoretical curves represent
the equilibrium overlap $m(\infty)$ and the basin of attraction $m_\alpha^c(0)$,
respectively.
It is seen that the theoretical results at second order are in reasonable 
agreement with numerical results.
We thus find that, in order to understand the correct behavior of the retrieval 
dynamics, we cannot ignore the effect of the temporal correlation of the noise.
As a result, it is shown that, even near saturation, the size of the basin of 
attraction remains nearly as large as that in the low loading rate regime
($\alpha\sim 0$).

In conclusion, we have proposed a generalized model of oscillator neural networks
to recall sparsely coded phase patterns in which some neurons are in a non-firing 
state and the other neurons encode information in the timings of spikes.
In the analysis of this model, we have found that, as in the standard binary model,
the storage capacity diverges as the activity level decreases.
At the same time, it is found that even near saturation, the basin of attraction 
remains sufficiently wide to allow for the recall of a memorized pattern from a 
noisy trigger pattern.
Hence, we can say that this system exhibits good associative memory.
Our results suggest the potential for information processing using the 
timings of neuronal firings,
though it remains to determine whether the results drawn for this model
hold more generally.

\acknowledgments
We would like to thank M. Okada and K. Kitano for helpful discussions.
This work is supported by the Japanese Grant-in-Aid for Science Research Fund from 
the Ministry of Education, Science and Culture.

%

\end{document}